\documentclass[twocolumn,prb,10pt,showpacs,superscriptaddress]{revtex4-1}
\usepackage[utf8]{inputenc}
\usepackage{graphicx}
\usepackage{amsmath}
\usepackage{amssymb}
\usepackage{bm}
\usepackage{hyperref}
\usepackage{color}
\usepackage{braket}
\usepackage{dsfont}

\begin{document}

\title{Topological Phase Detection in Rashba Nanowires with a Quantum Dot}

\author{Denis Chevallier}
\affiliation{Department of Physics, University of Basel, Klingelbergstrasse 82, 4056 Basel, Switzerland}
\author{Pawe\l{} Szumniak}
\affiliation{Department of Physics, University of Basel, Klingelbergstrasse 82, 4056 Basel, Switzerland}
\affiliation{AGH University of Science and Technology, Faculty of
Physics and Applied Computer Science,\\
al. Mickiewicza 30, 30-059 Krak\'ow, Poland}
\author{Silas Hoffman}
\affiliation{Department of Physics, University of Basel, Klingelbergstrasse 82, 4056 Basel, Switzerland}
\author{Daniel Loss}
\affiliation{Department of Physics, University of Basel, Klingelbergstrasse 82, 4056 Basel, Switzerland}
\author{Jelena Klinovaja}
\affiliation{Department of Physics, University of Basel, Klingelbergstrasse 82, 4056 Basel, Switzerland}

\date{\today}
\begin{abstract}
We  study theoretically the detection of the topological phase transition occurring in Rashba nanowires with proximity-induced superconductivity using a quantum dot. The  bulk states lowest in energy of such a nanowire have a spin polarization parallel or antiparallel to the applied magnetic field in the topological or trivial phase, respectively. We show that this property can be probed by the  quantum dot created at the end of the nanowire by external gates.  By tuning one of the two spin-split levels of the quantum dot to be in  resonance with nanowire bulk states one can detect the spin polarization of the lowest band via transport measurement. This allows one to determine the topological  phase of the Rashba nanowire independently of the presence of   Majorana bound states. 
\end{abstract}

\maketitle

\section{Introduction} Due to their possible applications for topological quantum computation~\cite{Alicea,Kitaev},  topological phases are one of the most studied topics currently in condensed matter physics. Such phases appear in various systems but most studies  focus on localized zero-energy modes, Majorana bound states (MBSs) \cite{Fu,MF_Sato,MF_Sarma,MF_Oreg,alicea_majoranas_2010,potter_majoranas_2011,
Klinovaja_CNT,Pascal,Bena_MF,halperin,gloria,manisha,Rotating_field,Ramon,Ali,trauzettel,
RKKY_Basel,RKKY_Simon,RKKY_Franz,Pientka,Ojanen,Yazdani,Franke,Meyer,
weithofer2014electron,thakurathi2015majorana,dmytruk2015cavity,manisha2,Fra,
reeg2014zero,thakurathi2015majorana,scharf2015probing,klinovaja2013giant}. One of the most promising systems are semiconducting Rashba nanowires (NWs) brought into proximity with an $s$-wave superconductor and subjected to an external magnetic field. Over the last years such systems were extensively studied experimentally \cite{Mourik,Marcus,Heiblum,Us,Marcus2,Frolov}. It is common to tune between the trivial and topological phase by changing external parameters such as the chemical potential or the  magnetic field. Experimentally, the presence of the topological phase is generally probed in transport setups by searching for a zero bias conductance peak generated by the MBSs. Unfortunately, this peak is far from being an unambiguous signature of a MBS and can come from other phenomena such as Andreev bound states, weak antilocalization, disorder or Kondo resonances \cite{Pikulinetal,PotterLee,Kellsetal,Diego,Sasakietal,
AkhmerovBeenakkerZBCP,LeeJiangHouzet,Nilssonetal,Franceschi}. It has been shown recently that the bulk states of such systems also carry  signatures of the topological phase \cite{Szumniak,Sau,Annica}. Indeed, the spin polarization along the externally applied magnetic field depends on the topological phase of the system  \cite{Szumniak}. In particular, the spin projection of the lowest electron (hole) band  is negative (positive) in the trivial phase, whereas it is opposite in the topological phase. This signature is quite universal because it is  also present  in multisubband systems and is robust against any kinds of weak, static, or magnetic disorder~\cite{Szumniak}. 

\begin{figure}[t]
\includegraphics[width=7.5cm]{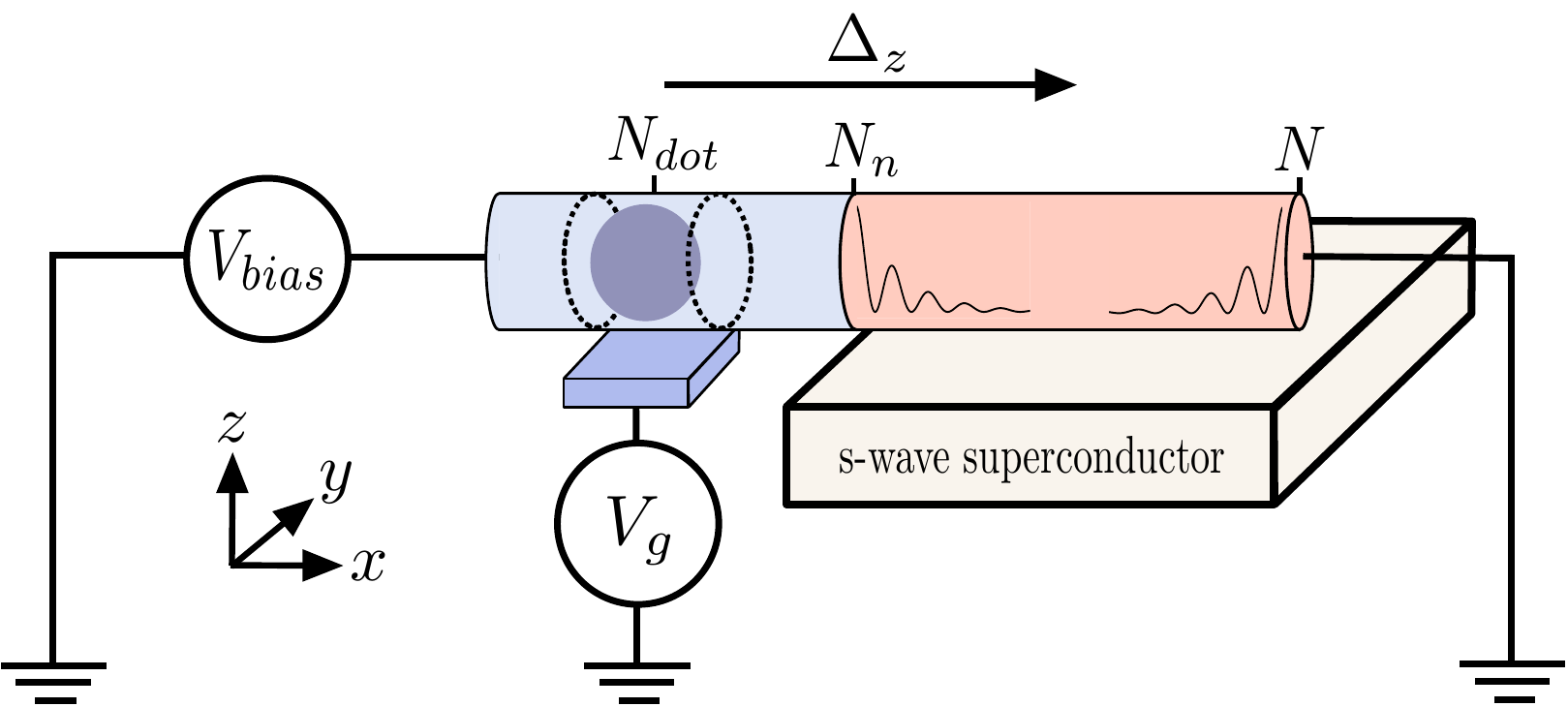}
\caption{The system consists of a semiconducting NW aligned along the $x$ axis with Rashba SOI and partially coupled to an $s$-wave superconductor  and subjected to an external magnetic field along the $x$ axis giving rise to a Zeeman energy $\Delta_z$. The part of the NW in contact with the superconductor (red part of the cylinder) is the TNW and can host MBSs (black curves) at its ends. The non-topological section of the NW (light blue part of the cylinder) is used to create a quantum dot (dark blue circle) by an external gate $V_g$. The NW is grounded and connected to a normal metal lead with a bias $V_{bias}$ allowing one to perform transport measurement.}
\label{fig:Fig1}
\end{figure}

In this work we focus on a detection scheme using a  quantum dot within the same Rashba NW (see Fig. \ref{fig:Fig1}) \cite{Carlos,Marcus2,Hoffman,Seridonio,Seridonio2,Szombati,Constantin,Sun,Seridonio3,pi,SanJose,Baranger,Flensberg,Avishai,Zheng,Leijnse,Rosenow}. 
The  proximity-induced superconductivity is induced only in one section of the NW that will be referred to as the topological nanowire (TNW) in the rest of the paper. The section of the NW not covered by the superconductor, referred to as the non-topological section, is used to create the quantum dot by external gates \cite{Marcus2}. The dot levels are spin-split by the external magnetic field applied along the NW.
Using a gate, we can move the dot levels and  align one of these spin-split states with the lowest in energy bulk band of the TNW that we aim to  probe. If the band and the dot level have the same spin polarization a current flows and, otherwise, not. By tuning the external parameters, we can tune between the trivial and topological phases of the TNW inducing a reversal of the spin polarization of the lowest bands, and, thus, switching on and off current through a particular dot level. Our central result is the differential conductance across the NW, obtained using Keldysh formalism and numerical evaluations, as a function of voltage bias and gate voltage on the quantum dot in the topological and trivial phases of the TNW.

The paper is organized as follows. In Sec. \ref{Model}, we present our model. In Sec. \ref{Protocol} and \ref{Conductance}, we explain the protocol of measurement and show the results for the differential conductance. Finally, we conclude in Sec. \ref{Conclusion}.

\section{Model}\label{Model} We consider a one-dimensional Rashba NW aligned along the $x$-axis brought partially into contact with an $s$-wave superconductor in presence of an external magnetic field applied in the $x$ direction, see  Fig. \ref{fig:Fig1}. The NW is divided into two sections. The TNW is coupled to the superconductor. The non-topological section hosts a quantum dot and is coupled via tunneling amplitudes $\hat{t}_L $ to a normal metal lead. By changing the applied bias voltage, $V_{bias}$, measured with respect to the chemical potential of the grounded wire, one induces an electrical current through the NW. The Hamiltonian of the total system $\tilde{H} = \tilde{H}_W + \tilde{H}_L + \tilde{H}_T(t)$, where the Hamiltonian describing the NW is written in the Nambu representation of the tight-binding model,
\begin{align}\label{substrate_hamiltonian}
\tilde{H}_W = \sum^N_{j=1} \tilde{\psi}^\dagger_j &\left[-\mu_j\tau_z+\Delta_{s,j} \tau_x+\Delta_z \sigma_x\right]\tilde{\psi}_j\notag\\
&+\sum^{N-1}_{j=1} \tilde{\psi}^\dagger_{j+1}\left[-\tilde{t}-i\tilde{\alpha}_j \sigma_y\right]\tau_z\tilde{\psi}_j +H.c. ,
\end{align}
with 
$\tilde{\psi}_j= (\psi^{\dagger}_{j,\uparrow}, \psi^{\dagger}_{j,\downarrow},  \psi_{j,\downarrow} , -\psi_{j,\uparrow})$. 
The operator $\psi^\dagger_{j\sigma}$ creates an electron with spin $\sigma$ at site $j$ of the chain with  $N$ sites.  The Pauli matrices $\sigma_\mu$ ($\tau_\mu$), $\mu=x,y,z$, act in spin (particle-hole) space. Here, $\tilde{t}$  is an effective hopping amplitude, $\Delta_z$ is the Zeeman energy, and $\tilde{\alpha}_j$ sets the strength of spin orbit interaction (SOI). In order to model a realistic setup, we choose different strengths of $\tilde{\alpha}_j$ for the non-topological section ($\tilde{\alpha}_{nt}$) and for the TNW ($\tilde{\alpha}_{t}$) as the superconductor attached to the TNW is believed to strongly enhance the SOI strength \cite{SOI_nonuniform}. The chemical potential, $\mu_j$, is defined to be $\mu_{t}$ for $j\geq N_n$ ({\it i.e.} in the TNW), $\mu_{nt}$ for $j<N_{dot}-L_d/2$ and $N_{dot}+L_d/2<j< N_n$ ({\it i.e.} in the non-topological section of NW {\it excluding} the dot), and $\mu_{dot}$ for $N_{dot}-L_d/2\leq j\leq N_{dot}+L_d/2$ which defines the quantum dot. Here, the center of the quantum dot of size $L_{d}$ is at position $N_{dot}$ and the chemical potential $\mu_{dot}$ is controlled by the external gate $V_g$. For convenience, we have chosen a step function in the chemical potential to create the dot. We have checked that the shape of the confinement potential does not matter for the results discussed below. We measure energy in units of the effective hopping, $\tilde{t}=1$. The superconducting pairing amplitude $\Delta_{s,j}$ is set to zero ($\Delta_s$) in the non-topological section (TNW). The bare retarded Green function encoding the properties of the nanowire reads in frequency space $\tilde{G}^{R}_{0}(\omega)=(\omega+i\delta -\tilde{H}_W)^{-1}$, with $\delta>0$ an infinitesimal needed to invert the matrix properly.
 
The normal metallic lead is described by the Hamiltonian $\tilde{H}_{L}=\sum_{k,\sigma} \xi_k \Psi^\dagger_{k,\sigma}\Psi_{k,\sigma}$, with $\xi_k  = k^2/2m-\mu_s$ and $\Psi_{k,\sigma}$ being the annihilation operator of an electron in the lead with spin $\sigma$ and momentum $k$. The tunneling Hamiltonian between the lead and NW is written as $\tilde{H}_{T}(t)=\sum_{k}\tilde{\Psi}^\dagger_{k} \tilde{t}_L(t) \tilde{\psi}_{j=1}+\textrm{H.c.}$, where $\tilde{\Psi}^\dagger_{k}$ ($\tilde{\psi}_{j=1}$) corresponds to the Nambu spinor composed of electron operators of the lead (of the left end of the NW) and $t$ denotes the time.
 The voltage difference between the lead and the substrate is included in the tunneling parameter via a Peierls substitution
 $\tilde{t}_L(t)=\hat{t}_L\tau_z e^{i \tau_z V_{bias} t}$. The total Green function of the system in the Nambu-Keldysh space can be expressed in frequency domain as
$\hat{G}^{-1}(\omega)=\hat{G}^{-1}_{0}(\omega)-\hat{\Sigma}(\omega)$, where $\hat{G}_{0}$ is the Green function of the NW and $\hat{\Sigma}(\omega)$ is the self-energy of the lead encoding all its properties as well as the tunneling rate between the lead and the NW, $\Gamma_L=\pi\nu_F\vert\hat{t}_L\vert^2$, where $\nu_F$ is the density of states per spin of the lead at the Fermi energy of the lead. By calculating the partition function in the Keldysh formalism (see Appendix A), we can extract the current flowing through the whole system,
\begin{align}
I_{c}=\frac{e}{2\hbar}\textrm{Tr}\{\tau_{z}\int\limits_{-\infty}^{+\infty}\frac{d\omega}{2\pi}
\textrm{Re}[\tilde{G}^{R}(\omega)\tilde{\Sigma}^{K}(\omega)+\tilde{G}^{K}(\omega)\tilde{\Sigma}^{A}(\omega)]\},
\end{align}
where $K$ and $A$ stand for the Keldysh and advanced component of the Green function and of the self-energy in the Keldysh formalism \cite{Zazunov,Chevallier,Chevallier3,LevyYeyati}.

\begin{figure}[bt]
\includegraphics[width=7.5cm]{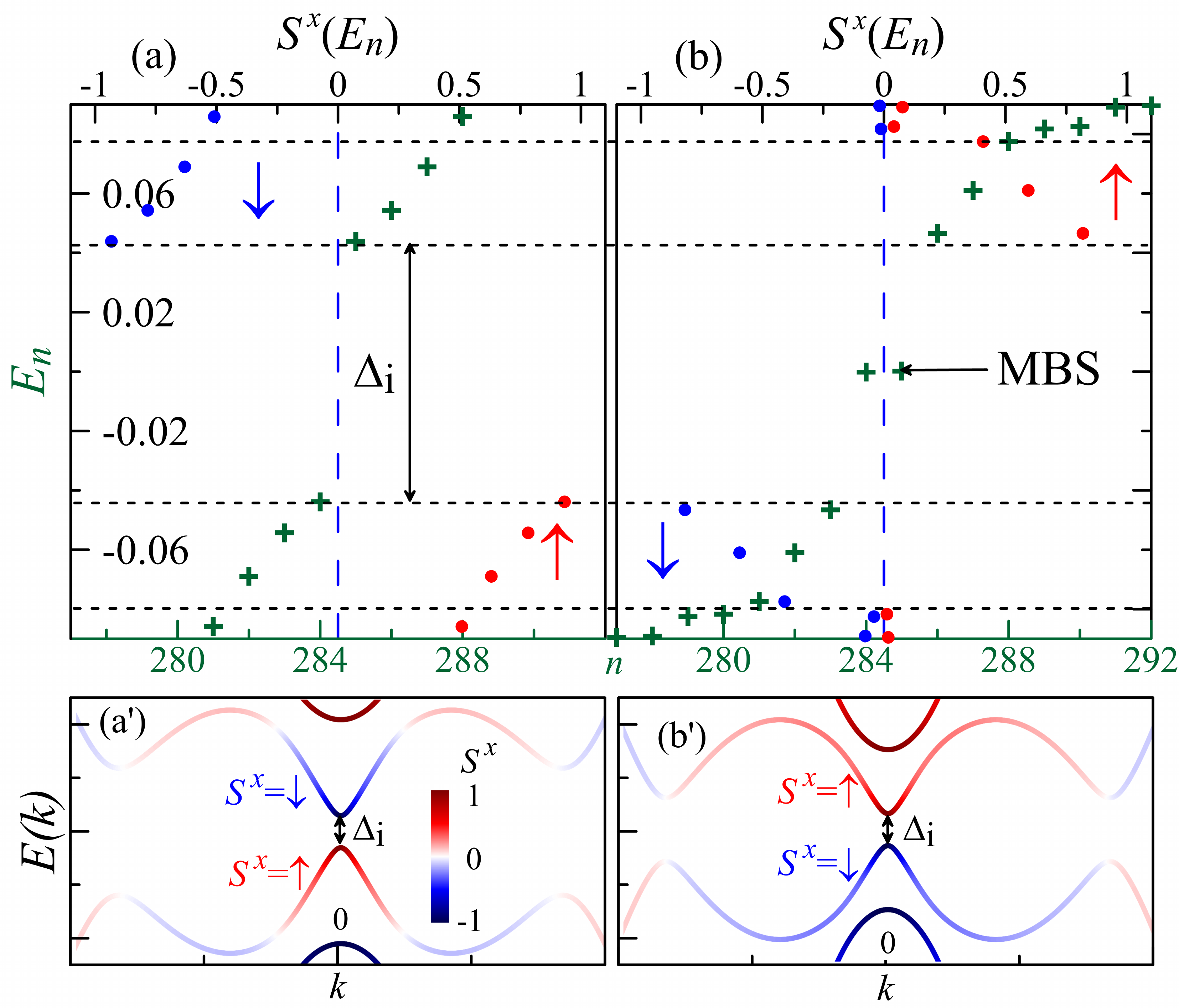}
\caption{Energy spectrum of the TNW (green crosses) and the spin polarization $S^x$ (blue and red dots) found in the tight binding model with $N=142$ in (a)  the trivial and (b) the topological phase. The corresponding quantities in the continuum limit are plotted in panels (a') and (b') for the trivial and the topological phase, respectively. The red and blue colors stand for the spin up and down polarization along the $x$-axis, respectively. For both models, we can clearly see the reversal of the spin polarization of the lowest bulk band as one goes through the topological phase transition. The parameters are chosen as follows: $\mu_{t}=-2$, $\tilde{\alpha}_t=0.5$, $\Delta_z=0.12$. We keep the topological gap $\Delta_i$ to be the same in the topological ($\Delta_s=0.08$) and trivial ($\Delta_s=0.16$) phases. 
This choice of parameters corresponds to $\Delta_{s,z}$  being on the order of 0.1-0.2 meV, while the SOI energy is around 0.2-0.3 meV.
}
\label{fig:Fig2}
\end{figure}

\section{Measurement protocol}\label{Protocol}
The spin polarization along the applied magnetic field of the lowest  energy bands of the TNW carries information about the topological phase transition \cite{Szumniak}, where the spin polarization of a given eigenstate is defined as ${\bf S}_{n}~=~\sum_{j=1}^N\Phi_n^\dagger(j) {\boldsymbol \sigma}\, \Phi_n (j)$ with $\Phi_n$ the $n^{th}$-eigenvector of the TNW with energy $E_n$. The spin polarization can be easily computed numerically from $\tilde{H}_W$ [see Figs. \ref{fig:Fig2}~(a) and (b)]  or analytically from the corresponding continuum model [see Figs. \ref{fig:Fig2}~(a') and (b')]. Independent of the approach, we clearly see the reversal of the spin polarization of the lowest bands around $k=0$ as the system goes through the topological phase transition, close to which the topological gap $\Delta_i=\vert\Delta_z-\Delta_s\vert$ is the smallest gap in the system \cite{Composite}.

\begin{figure}[t]
\includegraphics[width=7.5cm,height=4.7cm]{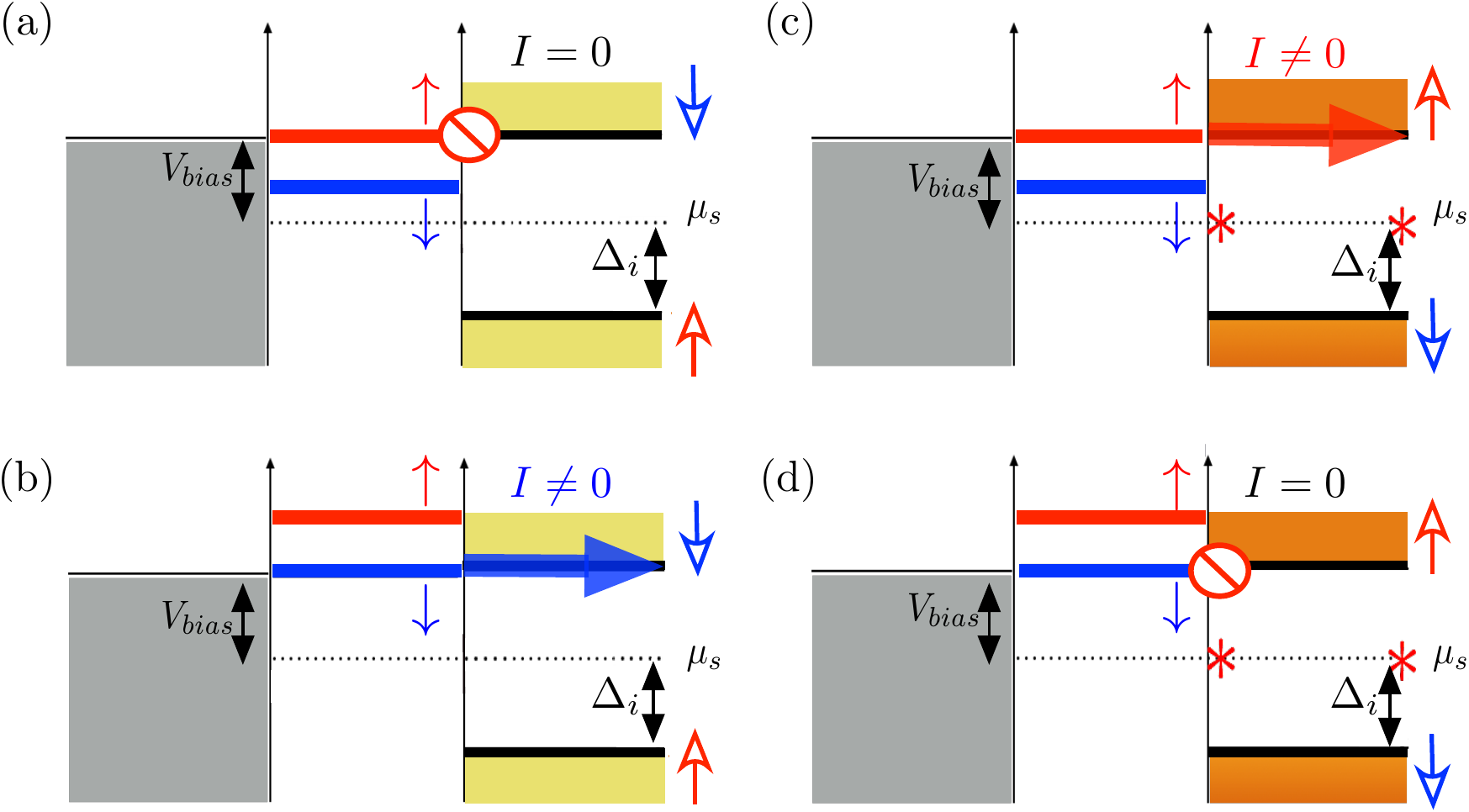}
\caption{Measurement protocol. The left part (gray) corresponds to the normal metallic lead, the middle part corresponds to the quantum dot with the two levels
representing spin-up (red) and spin-down (blue) states, and the right part corresponds to the TNW either [light yellow, (a) and (b)] in  the trivial or [orange, (c) and (d)] in the topological phase. In the latter case, there is a MBS (red star) at each end. We work in the regime close to the topological phase transition, so the topological gap $\Delta_i$ is  the smallest gap of the system. Note that, for simplicity, only the electron dot levels are drawn here. In the Nambu basis two hole levels are exactly at opposite energies. By tuning the levels of the dot by $V_g$, one can align them with the lowest bulk states of the TNW  and thus probe the spin polarization of these states. If the spin polarization of the dot level (small arrows) and of the bulk state (big arrows) are the same, there is a current $I$ flowing through the system for finite bias $V_{bias}$, which can be detected as a peak in the differential conductance. If their spin polarizations are opposite, the current is blocked, $I=0$.}
\label{fig:Fig3}
\end{figure}

The main goal of this work is to show how to detect this reversal of the spin polarization and, thus, the transition from trivial to topological phase of the TNW using a spin-split quantum dot within the same nanowire. In Fig. \ref{fig:Fig3}, we represent schematically the measurement protocol. The position of the chemical potential of the normal metallic lead is governed by the bias voltage $V_{bias}$, measured with respect to the reference  potential $\mu_s$ of the TNW. The two   levels (spin up and spin down) of the quantum dot are tuned by the gate voltage $V_g$ inside the topological gap $\Delta_i$ of the TNW, which can be either in a topological or trivial phase. In both phases, one should stay close to the topological phase transition such that the topological gap $\Delta_i$ is the smallest gap in the TNW. We note that the magnetic field controls both the topological phase of the TNW and the splitting of quantum dot levels. Fortunately, we are also able to tune the splitting of the dot levels by changing the length $L_{dot}$ of the quantum dot along the NW. Indeed, if the quantum dot is much larger than the SOI length $\lambda_{so}$, the Zeeman energy on the dot is strongly suppressed \cite{Mircea}. In the opposite limit $L_{dot} \lesssim \lambda_{so}$, the Zeeman energy starts to dominate and can already substantially split the two spin levels on the dot. By choosing a proper dot size, we can reach the configurations shown in Fig. \ref{fig:Fig3} with two dot levels inside the bulk gap of the TNW.

\begin{figure}[t]
\includegraphics[width=9.5cm]{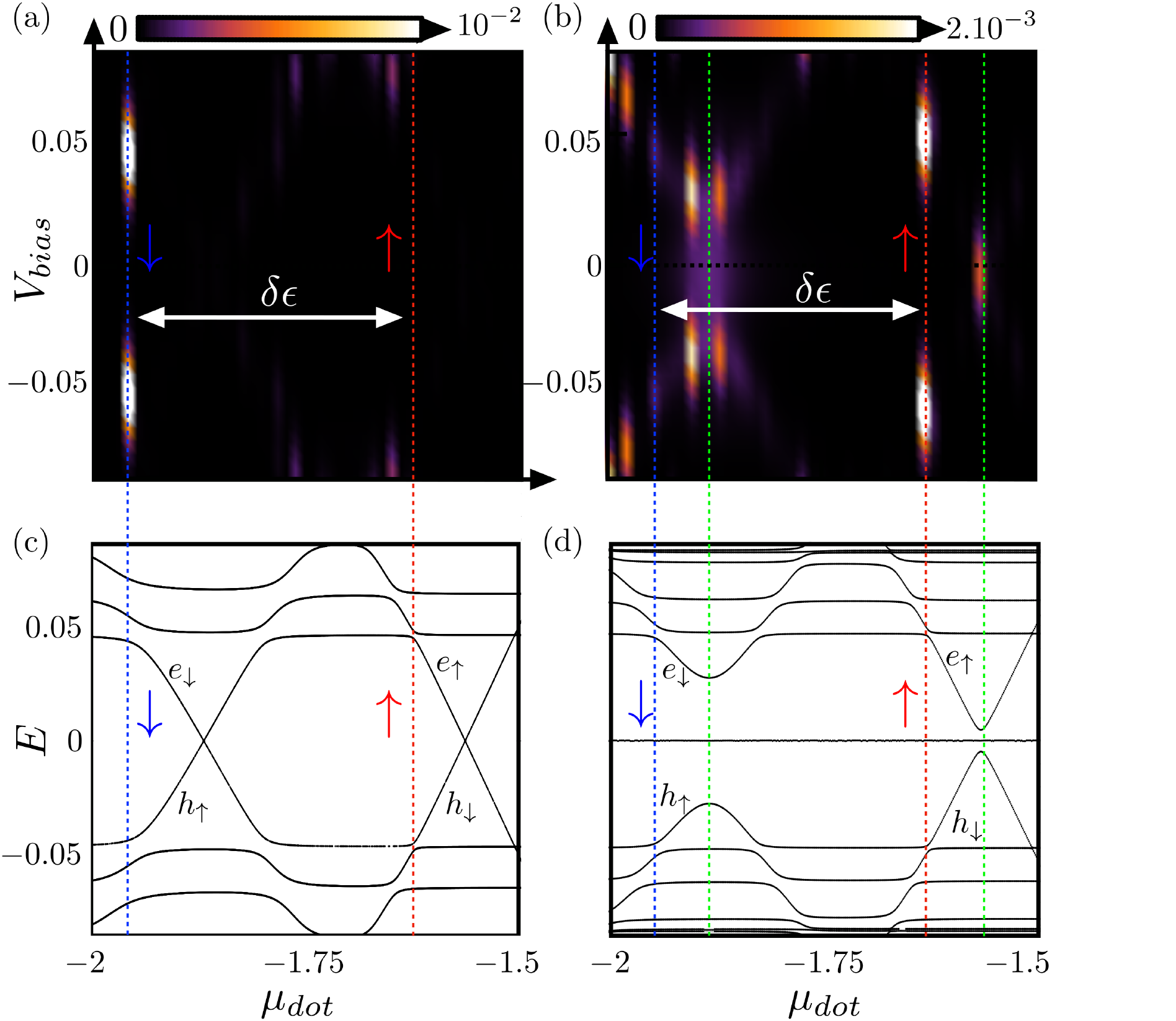}
\caption{Differential conductance $dI_c/dV_{bias}$ as a function of $\mu_{dot}$ and the bias $V_{bias}$ (in units of $e^2/h$) when the TNW is (a) in the trivial phase with $\Delta_s=0.16$ and (b) in the topological phase with $\Delta_s=0.08$ (these configurations correspond to the ones presented in Fig. \ref{fig:Fig2}). The corresponding energy spectrum of the system consisting of the dot and the TNW is shown in panels (c) and (d). The blue (red) line corresponds to the configuration in which the spin down (up) level of the dot is aligned with the lowest electron band. The green line correspond to the alignment of dot levels with the MBSs. Indeed, peaks in $dI_c/dV_{bias}$ appear if the bulk and dot levels of the same spin polarization are aligned, which allows one to distinguish between  topological and trivial phases. In both phases, the same shift of chemical potential on the dot, $\delta\epsilon\approx0.3$, is required to tune the setup from  the configuration in which the bulk states (or the MBS) are aligned in energy with the spin-up dot level to one in which they are aligned with the spin-down dot level. We note that in the topological phase [(b)], $\delta\epsilon$ can be read-out  as energy-distance between two near zero-energy MBS resonances (distance between green lines). Subsequently, this value of $\delta\epsilon$ can be used to determine the position of the missing peak at the edge of the bulk bands ($V_{bias}\approx \pm \Delta_i$) in the $dI_c/dV_{bias}$ signal (distance between blue and red lines). 
	The parameters of the system are: $\Delta_z=0.12$, $N=150$, ${N_n=9}$, $N_{dot}=5$, $L_{d}=3$, $\tilde{\alpha}_{nt}=0.01$, $\tilde{\alpha}_t=0.5$, $\mu_t=-2$, $\mu_{nt}=-2.37$, $\Gamma_L=0.02$, and $k_BT=1/200$ \cite{Comment}. 
}
\label{fig:Fig4}
\end{figure}

The principle of the measurement is straightforward: the gate $V_g$ allows us to push up or down the spin levels of the dot. When one dot level is energetically aligned with the lowest electron band of the TNW,  the electrons can enter and a current flows through the system [see Fig. \ref{fig:Fig3} (b) and (c)], provided the spin polarization of the dot level and the band are the same. However, if the spin polarization of the dot level is opposite to the one of the band [see Figs. \ref{fig:Fig3} (a) and (d)], the electrons cannot enter in the TNW leading to no contributions for the transport current. Therefore, if the system is in the trivial (topological) phase, the current flowing through the spin-down (spin-up) dot level should be finite and the current flowing through the  spin-up (spin-down) dot level will be strongly suppressed.

\section{Signal in differential conductance.} \label{Conductance}

\begin{figure}[t]
\includegraphics[width=9.5cm]{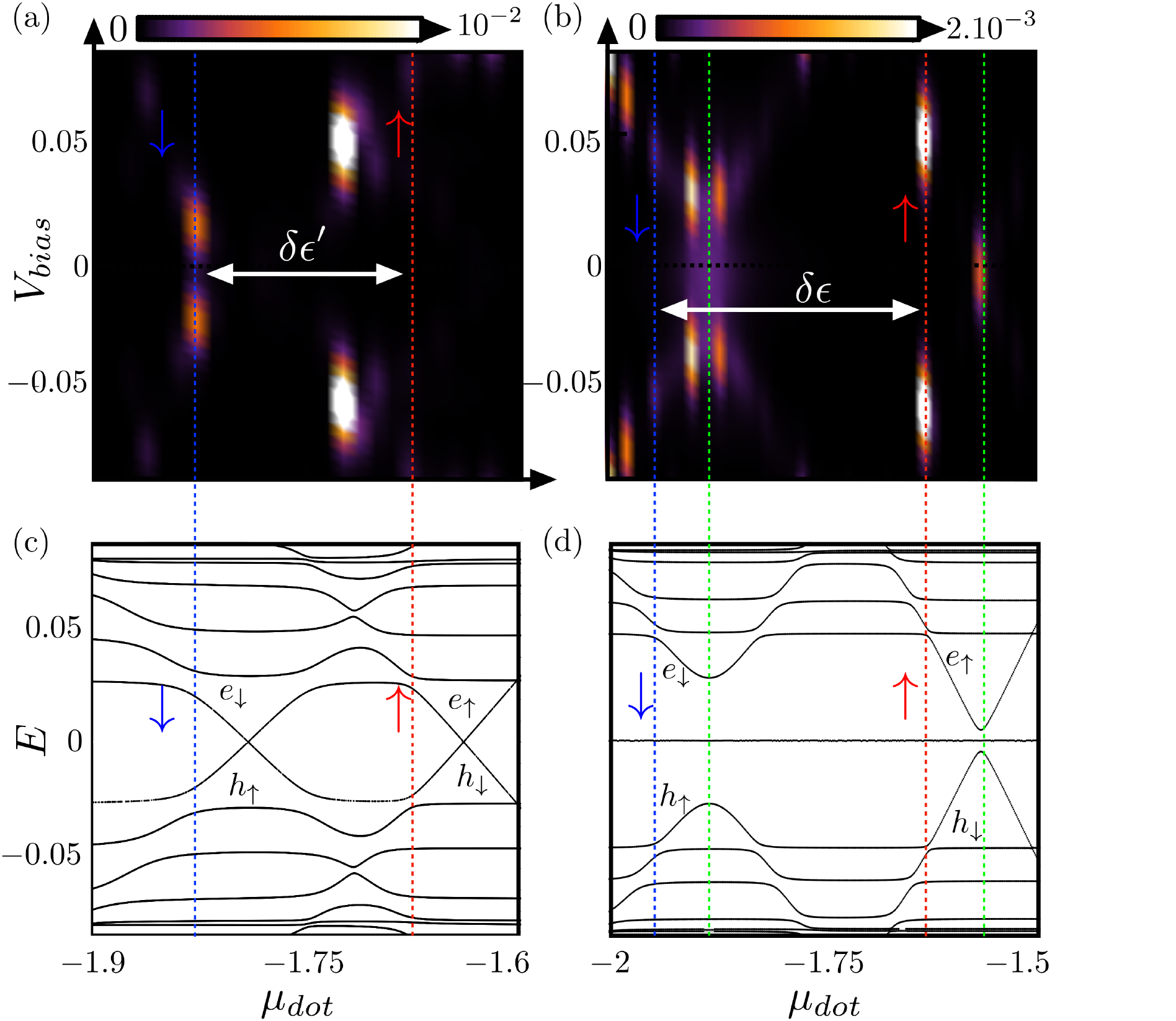}
\caption{The same as in Fig. \ref{fig:Fig4} except that $\Delta_s=0.08$ is kept constant and the topological phase transition is reached by tuning the magnetic field such that (a) $\Delta_z=0.06$ in the trivial phase and (b) $\Delta_z=0.12$ in the topological phase. As a result, $\delta\epsilon'<\delta\epsilon$. Again, the reversal of spin polarization can be detected in transport measurements.}
\label{fig:Fig5}
\end{figure}

Next, we confirm by numerical simulations that the topological phase transition can be detected in transport measurements.  As an example, we drive the system through the topological phase transition by changing the superconducting pairing amplitude such that the splitting of the dot levels stays the same. In Fig. \ref{fig:Fig4}, we plot the differential conductance $dI_c/dV_{bias}$ as a function of the chemical potential of the dot, $\mu_{dot}$, and the bias in the lead, $V_{bias}$. For convenience, we also show the corresponding band structure as a function of $\mu_{dot}$ in order to demonstrate that the features in the differential conductance correspond exactly to the point where the dot levels are tuned to be aligned with the lowest  TNW electron bands of the same spin polarization. Experimentally, the superconducting pairing amplitude is constant and one tunes the Zeeman field to reach the topological phase, see Fig.~\ref{fig:Fig5}.  In this case, by changing the magnetic field, one also changes the splitting between the dot levels meaning that the $\delta\epsilon'$ (trivial phase) is much smaller than $\delta\epsilon$ (topological phase). We find similar features as before, see  Fig.~\ref{fig:Fig5}, which clearly show the differences in the differential conductance between topological and trivial phase. However, for very small values of external field, extra features in the gap may appear due to crossing of different dot levels (see Appendix B). It is important to note that the parameters in Figs.  \ref{fig:Fig4} and  \ref{fig:Fig5} are in the experimental regime \cite{Marcus2}. 

\begin{figure}[t]
\includegraphics[width=6cm]{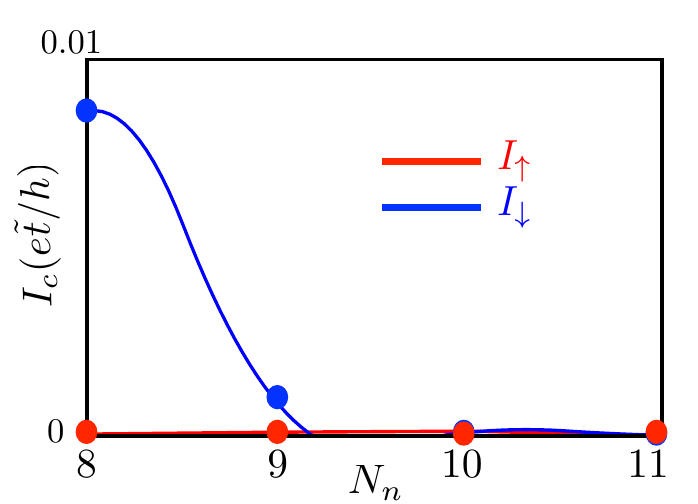}
\caption{ The current $I_c$ as a function of the distance $N_n-N_{dot}$ between the center of the dot $N_{dot}$ (kept constant) and the TNW $N_n$. The system is in the trivial phase and the parameters are the same as in Fig. \ref{fig:Fig4}(a). The bias voltage is fixed at the lowest electron bulk level of the TNW, $V_{bias}=0.05$. The gate voltage $V_g$ is tuned such that either the spin-up dot-level with the corresponding current $I_\uparrow$ (red line, $\mu_{dot}=-1.64$) or spin-down dot-level with the corresponding current $I_\downarrow$ (blue line, $\mu_{dot}=-1.94$) matches the lowest electron bulk level. Even if due to finite SOI, the current through dot levels with opposite spin-polarization is not exactly zero, the contrast between the two currents is substantial, {\it i.e.}, $I_\downarrow \gg I_\uparrow$.}
\label{fig:Fig6}
\end{figure}

The strength of the current depends on the effective tunneling between the dot and the TNW, and, thus, on the distance between them. Generally, the effective tunneling is given by the overlap of their wavefunctions. Due to the presence of SOI, the local spin polarization rotates in the $xz$-plane as a function of the position $x$ and, in principle, can affect our detection scheme \cite{Hoffman}. 
We have checked that the signal we get is mainly due to the spin polarization of the band and not due to an effective spin filtering coming from the rotation of the polarization axis. In Fig. \ref{fig:Fig6}, the system is in the trivial phase. The current through the spin-up level of the dot $I_\uparrow$ stays negligibly small no matter what  the distance is between the dot and the TNW. The current through the spin-down level $I_\downarrow$ is always finite and shows an exponential decay as the distance is increased. We note that there is no oscillatory behavior of the current, thus, the main signal is coming from the spin polarization of the bulk bands. The contrast between currents through two oppositely spin-split dot levels is substantial enough to use them as a detector of spin polarization and, thus, of the topological phase transition in the TNW. Finally, we note that by changing the strength of the magnetic field, the overlap between the dot and the TNW wavefunctions also changes, which affects the current.

\section{Conclusion}\label{Conclusion} 

We have shown that the topological phase of a TNW can be detected by measuring the current flowing between a spin-split quantum dot level and the lowest energy band of the TNW. The spin polarization of the lowest bands of the TNW reverses as the TNW is driven through the topological phase transition. As a result, the dot level serves as a spin filter and the current through spin-up (-down) level is finite only in the topological (trivial) phase, providing a clear experimental signature that can serve as an alternative way to detect the topological phase in TNWs independent of MBSs. Finally we note that a quantum dot is just a particular realization of a spin-probe to detect the bulk spin inversion due to the topological phase transition; alternatively, the same goal can be achieved by making use of spin-polarized STM tips \cite{TD_SRI, Yazdani_STM_TIP}.

\emph{Acknowledgments.} We acknowledge helpful discussions with M. Deng, A. Baumgartner, C. Juenger, and P. Makk. This work was supported by the Swiss National Science Foundation and  NCCR QSIT.

\appendix

\section{Details on the current calculation in the Keldysh formalism}

The voltage difference between the tip and the substrate is included in the tunneling amplitude via a Peierls substitution, $\tilde{t}_L(t)=\hat{t}_L\tau_z e^{i \tau_z V_{bias} t}$.
The bare Green function encoding the properties of the nanowire reads
$\tilde{G}_{0}^{ss'}(t,t')=-i\langle T_{C}\{\tilde{\psi}^s(t)\tilde{\psi}^{s'\dagger }(t')\}\rangle_0,$
where $T_C$ is the time ordering operator along the Keldysh contour with $s,s'$ labeling the branches and $\tilde{\psi}=(\tilde{\psi}_{j=1},...,\tilde{\psi}_{j=N})$. The total Green function of the system
reads
$\tilde{G}^{ss'}(t,t')=-i\langle T_{C} \{S(\infty)\tilde{\psi}^s(t)\tilde{\psi}^{ s'\dagger}(t')\}\rangle_0$,  
where the evolution operator along the contour
$S(\infty)=T_{C}\exp\{-i\int_{-\infty}^{+\infty}\!\!dt\,\sum\limits_{s=+,-}\eta_{z}^{ss}\tilde{H}_{T}^{s}(t)\},$
and $\eta_{z}$ is the $z$-Pauli matrix in the Keldysh space. Because the lead degrees of freedom are quadratic in $\tilde{H}$, the evolution operator can be averaged over it, 
\begin{equation}
\left\langle 
S(\infty)\right\rangle_{leads}=T_{C}\exp\left[-i\int_{C}\!\!dt_{1}dt_{2}\,
\hat{\psi}^{\dagger}(t_{1})\hat{\Sigma}_{L}(t_{1},t_{2})\hat{\psi}(t_{2})\right],
\end{equation}
where we introduce the spinor $\hat{\psi}$ in the Nambu-Keldysh space. The self-energy associated with the lead can be written as
$[\hat{\Sigma}_{L}(t_{1},t_{2})]_{ii}=\hat{\Sigma}_{i,L}(t_{1},t_{2})$, where all the components are zero except at the site $i=1$ where the lead is attached to the NW,
\begin{equation}\label{sigma_time}
\hat{\Sigma}_{i,L}(t_1,t_2)=(\tilde{t}_{L}^{\dagger}(t_1)\otimes\eta_{z})\hat{g}_{L}(t_{1}-t_{2})(\eta_{z}\otimes\tilde{t}_{L}(t_{2}))~.
\end{equation}
Here, $\hat{\Sigma}_L$, $\hat{g}_{L}$ are matrices in Nambu-Keldysh space,
with $\hat{g}_{L}(t-t^{\prime})$ the Green function of electrons in the lead. In the literature, they are typically given in the frequency domain: $\tilde{g}^{R/A}_L(\omega)=\mp i \pi \nu_F$ and $\tilde{g}^K(\omega)=(1-2f(\omega))(\tilde{g}^{R}_L(\omega)-\tilde{g}^{A}_L(\omega))$. The superscripts $R,A,K$ correspond to the components in the rotated Keldysh space. The self-energy in frequency domain can be calculated easily by inserting these functions into Eq.~(\ref{sigma_time}) and performing a Fourier transform leading to
\begin{eqnarray}
&&\tilde{\Sigma}_{L}^{A/R}(\omega)=\pm{}i\Gamma_L,\\
&&\tilde{\Sigma}_{L}^{K}(\omega)=-2i\Gamma_L
\begin{pmatrix}
  {\rm{}tanh}(\beta\omega_-/2) & 0 \\
  0 & {\rm{}tanh}(\beta\omega_+/2)\notag
\end{pmatrix},
\end{eqnarray}
where $\omega_{\pm}=\omega\pm V_{bias}$ and $\Gamma_L=\pi\nu_F\left|\hat{t}_L\right|^2$ is the tunneling rate between the NW and the lead and $\beta=1/k_BT$ with $T$ the temperature of the electrons in the lead. The Green function $\tilde{G}$ remains to be determined. To do this, we write the Dyson equation in the frequency domain and we obtain the various components of $\tilde{G}$ in the rotated Keldysh space
\begin{align}\label{dyson_equation}
\tilde{G}^{R/A}(\omega)^{-1}&=\tilde{G}^{R/A}_{0}(\omega)^{-1}-\tilde{\Sigma}^{R/A}_{L}(\omega),\\
\tilde{G}^{K}(\omega)&=\tilde{G}^{R}(\omega)\tilde{\Sigma}^{K}_{L}(\omega)\tilde{G}^{A}(\omega),\label{dyson_equation2}
\end{align}
with $\tilde{G}^{R/A}_{0}(\omega)=(\omega\pm i\delta-\tilde{H}_{W})^{-1}$.
The current between the NW and the lead can be calculated via the change in the charge density 
$\frac{\partial \rho}{\partial t}=\frac{1}{i}[\rho,\tilde{H}]$ leading to
\begin{equation}
I(t)=\frac{i}{2}[\sum_{k}\tilde{\Psi}^\dagger_{k}\tau_z\tilde{\Psi}_{k},\tilde{H}_T(t)]=\frac{i}{2}\sum_{k}\tilde{\Psi}^\dagger_{k}\tau_z\tilde{t}_L(t)\tilde{\psi}_{j=1}.
\end{equation}
 To compute it, it is convenient to introduce counting fields $\gamma(t)$, which appear in the tunneling amplitudes as $\tilde{t}_L(t)\rightarrow \tilde{t}_L(t)e^{i\eta_z\otimes \tau_z \gamma(t)/2}$. The average current from the nanowire into the lead can then be calculated as the first derivative of the Keldysh partition function,
\begin{equation}
I_c=\left\langle I\right\rangle=i\frac{1}{Z[0]}\left.\frac{\delta Z\left[\gamma\right]}{\delta \gamma(t)}\right|_{\gamma=0},
\end{equation}
where $Z[\gamma]=\left\langle S(\infty,\gamma)\right\rangle_0$ and $S(\infty,\gamma)$ is the evolution operator in which the counting fields were introduced.  After performing the derivative and a Fourier transform to go to frequency domain, we can write the average current  in terms of the advanced, retarded, and Keldysh components by taking the trace over the Keldysh space and get
\begin{eqnarray}
I_c=\frac{e}{2\hbar}\textrm{Tr}\{\tau_{z}\int\limits_{-\infty}^{+\infty}\frac{d\omega}{2\pi}
\textrm{Re}[\tilde{G}^{R}(\omega)\tilde{\Sigma}^{K}(\omega)+\tilde{G}^{K}(\omega)\tilde{\Sigma}^{A}(\omega)]\}.
\end{eqnarray}

\section{$dI_c/dV_{bias}$ and Band structure for small magnetic field.}

\begin{figure}[ht]
\includegraphics[width=6cm]{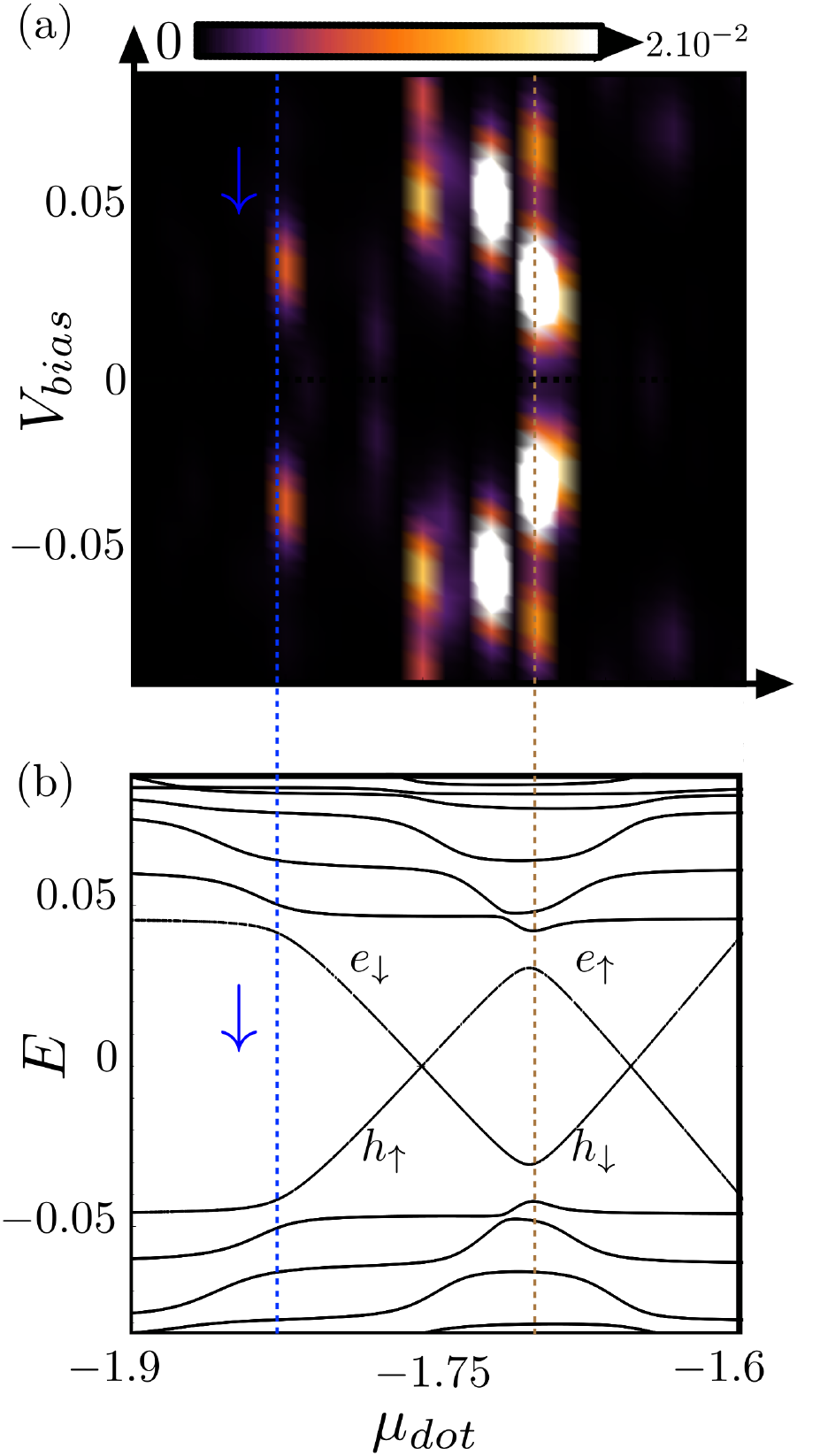}
\caption{ The same as in Fig. 5(a). The system is in the trivial phase with $\Delta_z=0.04$. The anticrossings between electron and hole levels of the quantum dot inside the bulk gap of the TNW lead to extra features inside the bulk gap (brown dashed line).}
\label{fig:Fig7}
\end{figure}

As mentioned in the main text, the Zeeman field not only plays an important role in tuning the TNW into the topological phase but it also sets the splitting between dot levels. In our study, we have noticed that for small magnetic field the dot levels can interact between themselves within the gap and give rise to extra features in the differential conductance inside the bulk gap, see Fig. \ref{fig:Fig7}. In this particular configuration, there is a crossing between electron and hole levels inside the bulk gap of the TNW.  As a result, one can clearly see an additional feature appearing in transport experiments (brown dashed line).

\begin{thebibliography}{99}
\bibitem{Comment} We have checked that our results are robust when we change the tunneling strength and the temperature. Indeed, the transport signature remains clearly visible for a reasonable range of $\Gamma_L$ that is  reachable experimentally (basically the condition $\Gamma_L\ll\delta\epsilon$ has to be fulfilled). The same holds for 
temperatures up to $k_BT\lesssim 1 /100$.
\end{thebibliography}
\end{document}